\documentclass[12pt]{iopart}
\usepackage{bbm}
\usepackage{graphicx}
\usepackage{fancyhdr, color}
\usepackage{hyperref}

\expandafter\let\csname equation*\endcsname\relax
\expandafter\let\csname endequation*\endcsname\relax
\usepackage{amsmath,amssymb} 
\usepackage{cite}

\newcommand{\op}[1]{{\sf #1}}

\newcommand{\oL}{\op{L}}
\newcommand{\oQ}{\op{Q}}
\newcommand{\oPhi}{\op{\Phi}}
\newcommand{\oH}{{\mathsf H}}

\newcommand{\TCM}{{T_{\rm CM}}}
\newcommand{\Rf}{{R_{\rm eff}}}
\newcommand{\nfb}{{\mathcal{V}_{\rm fb}}}

\newcommand{\ntot}{{\mathcal{V}_{\rm total}}}
\newcommand{\nR}{{\mathcal{V}_{\rm R}}}
\newcommand{\Tn}{{T_{\rm fb}^{\rm n}}}
\newcommand{\Tmin}{{T_{\rm fb}^{\rm min}}}

\begin{document}
\title{Levitated electromechanics: all-electrical cooling of charged nano- and micro-particles.}
\date{\today}
\author{Daniel Goldwater$^1$, Benjamin A. Stickler$^2$, Lukas Martinetz$^2$, Tracy E. Northup$^{3}$, Klaus Hornberger$^2$ and James Millen$^{4}$}
\address{$^1$Department of Physics and Astronomy, University College London, Gower Street, London WC1E 6BT, United Kingdom}
\address{$^2$University of Duisburg-Essen, Faculty of Physics, Lotharstra{\ss}e 1, 47048 Duisburg, Germany.}
\address{$^3$Institut f{\" u}r Experimentalphysik, Universit{\" a}t Innsbruck, Technikerstra{\ss}e 25, 6020 Innsbruck, Austria.}
\address{$^4$Department of Physics, King's College London, Strand, London, WC2R 2LS, UK.}

\begin{abstract}
We show how charged levitated nano- and micro-particles can be cooled by interfacing them with an $RLC$ circuit. All-electrical levitation and cooling is applicable to a wide range of particle sizes and materials, and will enable state-of-the-art force sensing within an electrically networked system. Exploring the cooling limits in the presence of realistic noise we find that the quantum regime of particle motion can be reached in cryogenic environments both for passive resistive cooling and for an active feedback scheme, paving the way to levitated quantum electromechanics.
\end{abstract}

\maketitle
\section{Introduction}

The control of nano- and micro-scale objects in vacuum is of great importance for a wide range of applications \cite{Maher2015}, from the study of individual proteins \cite{Bruce1994}, viruses \cite{Fuerstenau2001} and bacteria \cite{Peng2004}, to the simulation of astrophysical processes via the study of dusty plasmas \cite{Morfill2009}, to the detection of tiny forces \cite{Ranjit2015, Kuhn2017a}. It is predicted that the motion of levitated nanoparticles can be controlled even at the quantum level \cite{Chang2010, Barker2010, Isart2010,Stickler2016}, enabling studies of macroscopic quantum physics \cite{Arndt2014}, and the contribution of levitated high quality-factor  oscillators in quantum technologies as coherent storage or signal transduction devices.

To achieve quantum control of nano- and micro-particles, their motion must be cooled towards the quantum regime, which is viable via optical cavity cooling \cite{Chang2010, Barker2010, Isart2010, Kiesel2013, Asenbaum2013, Millen2015, Stickler2016, Kuhn2017b,Fonseca2016} or parametric feedback \cite{Gieseler2012,Jain2016}, achieving temperatures below $500\,\mu$K (a few tens of motional quanta). However, these optomechanical methods are limited by the very optical fields that are used for cooling, through optical absorption \cite{Millen2014, Hebestreit2017}, photon scattering \cite{Jain2016}, and instability at low pressures \cite{Kiesel2013, Millen2014, Ranjit2015, Vovrosh2017}.

Charged particles can instead be levitated in ion traps \cite{Paul1953, Paul1990}, with trapped atomic ions at the forefront of quantum information processing technology \cite{Duan2010, Singer2010}. Their motion can be detected and cooled via their coupling to the trap electrodes \cite{Wineland1975, Itano1995}, either resistively or using active feedback methods \cite{DUrso2003, Guise2010}. Extending these techniques to nano- and micro-particles will enable manipulation and control of massive charged particles by electronic circuitry. This opens the field of \emph{levitated electromechanics}, holding the promise of scalability and network-integration.

In this work, we present an all-electrical levitation, detection and cooling scheme for charged nano- and micro-particles, which will operate stably under vacuum conditions while avoiding optical scattering and absorption heating. Our simulations show that sub-Kelvin particle temperatures are achievable with room temperature circuitry via resistive and feedback-cooling. The applicable range of sizes, from sub-nanometre to several micrometres, and materials, including metallic clusters, dielectric particles and biological objects, illustrates the universality of this technology. Under ambient conditions, levitated electromechanics will offer a simple and robust pre-cooling method for optical cavity cooling, and state-of-the-art force sensing. This scheme is suited to cryogenic environments, where milli-Kelvin temperatures allow reaching the deep quantum regime. 

We first derive the equations-of-motion of a charged particle coupled to an $RLC$ circuit, and show how they can be quantized. Then we discuss the electronic detection of motion, followed by the potential to cool through passive-resistive and active-feedback methods. Finally we discuss the application of levitated electromechanics to displacement and force sensing.

\section{Equations of motion}

\begin{figure}[t]
	 {\includegraphics{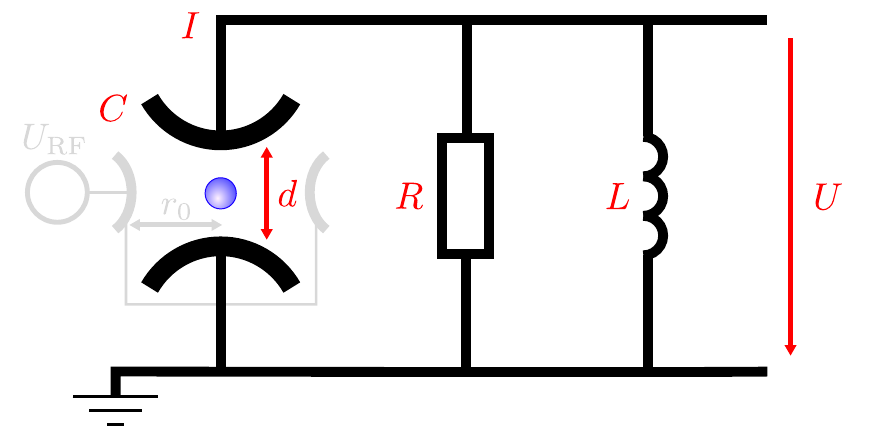}}	\centering
\caption{\label{fig:circ} 
Circuit diagram for the electronic detection of charged particle motion for an object levitated in a quadrupole ion trap (gray diagram). The motion of the trapped particle is picked-up by the endcap electrodes separated by $d$, and induces a current $I$ in the circuit. A voltage $U$ is generated across the resistance $R$ of the circuit. Resonant detection is possible by forming either a series or parallel $RLC$ circuit (with parallel illustrated here), where the capacitance $C$ is due to the trap electrodes plus any parasitic capacitance.}
\end{figure}

We consider a spherical particle of mass $M$ and charge $q$ levitated in a potential $V(z)$, with trapping frequency $\omega_z$, which is levitated between plates of separation $d$ and capacitance $C$. As illustrated in fig.~\ref{fig:circ} the two electrodes are joined via an inductance $L$ to form an $LC$ circuit. While in this work we envision that $V(z)$ is provided by a Paul trap, with the capacitor formed by the endcap electrodes, we note that the levitating potential could also be optical. The equations of motion for a charged particle in a Paul trap are provided in Appendix 1. The interaction with electrostatic mirror charges formed in the endcap electrodes is negligible for the trap parameters considered below.
 
The dynamics of the charged particle can be derived using the Shockley-Ramo theorem \cite{shockley1938,sirkis1966}. It relates the current $I$ flowing in the circuit with the particle momentum $p = M \dot{z}$ and the voltage drop $U$ across the capacitor. For now we consider the conservative $LC$ circuit neglecting resistance, yielding

\begin{equation} \label{eq:indI}
I = - \frac{q}{d} \frac{p}{M} + C \dot{U}.
\end{equation} 

We use the circuit-charge $Q$ on the capacitor as the generalized coordinate of the circuit, so that the magnetic flux through the inductor $\Phi = L \dot{Q}$ acts as its associated canonical momentum, and exploit Kirchhoff's law, $U = - \dot{\Phi}$, to obtain the canonical equations-of-motion,

\begin{subequations}\label{eq:circpart}
\begin{eqnarray} 
\dot{Q} & = & \frac{\Phi}{L} \label{eq:circa} \\
\dot{\Phi} & = & - \frac{Q}{C} - \frac{q}{C d} z. \label{eq:circb}
\end{eqnarray}

The first term in eqn.~\eqref{eq:circb} gives rise to a harmonic oscillation of frequency $\omega_{LC}=1/\sqrt{LC}$, while the second term accounts for the linear coupling to the particle position $z$.

The back-action of the circuit dynamics onto the particle motion can be derived by noting that the voltage offset $U$ across the capacitor induces a constant force $- q Q/C d$ acting on the particle. Neglecting charging effects due to the formation of mirror charges, which are on the order of $q^2 /
M C d^2 \ll \omega_z^2$, one thus obtains the equations-of-motion for the particle:

\begin{eqnarray}
\dot{z} & = & \frac{p}{M} \\
\dot{p} & = & - \partial_z V(z) - \frac{q}{Cd} Q.
\end{eqnarray}
\end{subequations}

Equations \eqref{eq:circpart} describe the coupled classical dynamics of a charged particle trapped inside an LC circuit. The Hamiltonian function associated with these dynamics can be identified,

\begin{equation} \label{eq:ham}
H = \frac{\Phi^2}{2L} + \frac{Q^2}{2 C} + \frac{p^2}{2 M} + V(z) + \frac{q}{C d} Q z.
\end{equation}
For harmonic $V(z)$ the particle-circuit system behaves as two linearly coupled harmonic oscillators.

\subsection{Dissipation} \label{sec:dissipation}

We now consider the role of a resistor, of resistance $R$ and temperature $T_R$, connected in series with the $LC$ circuit. Its action  is described by adding the damping and diffusion term $- \Gamma \Phi + \sqrt{2 \Gamma L k_{\rm B} T_R} \xi(t)$ to eqn.~\eqref{eq:circb}. Here, the damping rate of the circuit is $\Gamma = R/L$, and the second term with white noise $\xi(t)$, i.e.~$\langle \xi(t + \tau) \xi(t) \rangle = \delta(\tau)$, describes circuit thermalization at temperature $T$. Due to the linear coupling between the circuit and the particle (eqn.~\eqref{eq:ham}), the particle will thermalize with the circuit resistance at temperature $T_R$. The corresponding timescale $1/\gamma$ is determined by the circuit parameters $R$, $L$ and $C$ as well as by the particle charge $q$ and mass $M$. Dissipation in a parallel $RLC$ circuit can be described in a similar fashion by adding a friction and diffusion term, with parallel friction rate $\Gamma = 1/RC$, to the equation-of-motion for the circuit charge.

\subsubsection{Adiabatic damping}

If the circuit follows the particle motion adiabatically we can eliminate the former from the equations-of-motion in eqn.~\eqref{eq:circpart}. Neglecting diffusion for the moment, one obtains from eqn.~\eqref{eq:circa} and \eqref{eq:circb} the quasi-static expressions for a series $RLC$ circuit,

\begin{subequations}
\begin{eqnarray}
Q \simeq -\frac{q}{d}z + \frac{\Gamma L C q}{d M}p, \\
\Phi \simeq -\frac{q L}{ M d}p. 
\end{eqnarray}
\end{subequations}

These yield the non-conservative equations of motion for the particle,

\begin{subequations} \label{eq:partad}
\begin{eqnarray}
\dot{z} & = &\frac{p}{M} \\
\dot{p} & = & - \partial_z V(z) - \frac{\Gamma L q^2}{M d^2} p,
\end{eqnarray}
\end{subequations}
which allows us to identify the adiabatic damping rate

\begin{equation} \label{eq:addamp}
\gamma_{\rm ad} = \frac{\Gamma L q^2}{M d^2}.
\end{equation}

A similar calculation for the parallel RLC circuit yields $\gamma_{\rm ad} = 0$ since the capacitor is effectively shorted out in this scenario. In the derivation of eqns.~\eqref{eq:partad} we neglected the additional attractive force related to the appearance of electrostatic mirror charges. This contribution is negligible in the present case, and causes a fractional frequency shift of $q^2/MCd^2 \omega_z^2 \sim 10^{-6}$.

\subsubsection{Damping on resonance}

If the circuit is on resonance with the particle motion, $\omega_{LC} = \omega_z$, the oscillatory dynamics of the particle can effectively cancel the inductance, leading to a boosted friction rate. This can be made evident by Fourier-transforming the particle-circuit equations-of-motion for the series $RLC$ circuit and eliminating the circuit,
\begin{equation}
\omega^2 \widetilde{z}(\omega) = \left ( \omega_{z}^2 + \frac{q^2 \omega_z^2}{MCd^2} \frac{\omega^2 - \omega_z^2}{(\omega^2 - \omega_z^2)^2 + \omega^2 \Gamma^2} \right ) \widetilde{z}(\omega) + i \frac{q^2 \omega_z^2}{MCd^2} \frac{\omega \Gamma}{(\omega^2 - \omega_z^2)^2 + \omega^2 \Gamma^2} \widetilde{z}(\omega).
\end{equation}
For $\Gamma^2 \gg |\omega^2 - \omega_{z}^2 |$, this equation yields the on-resonance particle friction rate
\begin{equation} \label{eq:resdamp}
\gamma_{\rm res} = \frac{q^2}{MC \Gamma d^2},
\end{equation}
which increases with decreasing circuit dissipation. A similar calculation for the parallel $RLC$ circuit yields the same result for the friction rate. This resonant enhancement of friction is in agreement with the damping rate observed for trapped atomic ions and electrons \cite{Brown1986, Domizio2015}. Note that the series adiabatic friction rate \eqref{eq:addamp} is equal to the parallel on-resonance damping rate \eqref{eq:resdamp}.

\subsection{Quantum dynamics}

We consider the situation where particle and circuit are cooled to their ground-state or close to it. The coupled quantum dynamics of the combined particle-series $RLC$ circuit state $\rho$ can be modeled by the Linblad master equation \cite{breuer2002},
\begin{equation} \label{eq:master}
\partial_t \rho = - \frac{i}{\hbar} \left [ \oH, \rho \right ] - \frac{i \Gamma}{4 \hbar} \left [ \left \{\oQ, \oPhi \right \}, \rho \right ] + \frac{\Gamma}{2} \left ( \oL \rho \oL^\dagger - \frac{1}{2} \left \{ \oL^\dagger \oL, \rho\right \} \right ),
\end{equation}

\noindent where $\oH$ is the quantized Hamiltonian eqn.~\eqref{eq:ham} (operators are denoted by sans-serif characters), and the Lindblad operators are
\begin{equation}
\oL = \frac{\sqrt{4 L k_{\rm B} T_R}}{\hbar} \oQ + \frac{i}{\sqrt{4 L k_{\rm B} T_R}} \oPhi.
\end{equation}

The completely positive master equation \eqref{eq:master} describes the Markovian thermalization dynamics of the particle-circuit state $\rho$. For large circuit temperatures, where the term proportional to $1/T_R$ can be neglected, one obtains

\begin{equation}
\partial_t \rho = - \frac{i}{\hbar} \left [ \oH, \rho \right ] + \frac{\Gamma}{2 i \hbar} \left [ \oQ, \left \{\oPhi , \rho \right \} \right ] - \frac{\Gamma L k_{\rm B} T_R}{\hbar^2} \left [ \oQ, \left [ \oQ, \rho \right ] \right ].
\end{equation}

The second term on the right hand side describes damping of the circuit momentum $\Phi$ with friction rate $\Gamma$, while the third term describes charge diffusion. The corresponding diffusion constant $\Gamma L k_{\rm B} T_R$ is in accordance with the fluctuation-dissipation relation. If the trapping potential  is harmonic the master equation describes thermalization towards the Gibbs equilibrium $\rho_{\rm eq} = \exp ( - \oH/k_{\rm B} T_R ) / Z$, and the expectation values of the canonical phase-space observables exhibit the classical thermalization dynamics (eqns.~\eqref{eq:circpart}).

\section{Detection of the particle motion}

Experiments demonstrating trapping of single electrons \cite{Wineland1973, DUrso2005}, ions \cite{Dehmelt1986} and protons \cite{Guise2010} as well as electron \cite{Wineland1975} and proton \cite{Church1969} clouds, have noted that it is possible to detect the classical motion of the charged particle(s) via the image current $I$ that they generate in the endcap electrodes of Paul or Penning traps \cite{Brown1986}. We now show that this is also possible for spherical nano- and micro-scale particles.  Under the assumption that $\Gamma \gg \omega_z$ the circuit adiabatically follows the particle motion, and eqn.~\eqref{eq:indI} reduces in lowest order of the particle velocity to

\begin{equation} 
\label{eqn:current}
	I = -\frac{q\eta}{d}\dot{z},
\end{equation}

\noindent where we have introduced the geometrical factor $\eta$ to account for the shape of the pick-up electrodes, which we assign the realistic  value of $\eta=0.8$ for slightly parabolically shaped electrodes \cite{Itano1995}. The maximum velocity given by the equipartition theorem is $\dot{z}_{\mathrm{max}} = \sqrt{k_{\rm B}\TCM/M}$, where $\TCM$ is the centre-of-mass temperature of the particle. This implies a peak induced current of 

\begin{equation} 
\label{eqn:maxcurrent}
	I_{\mathrm{max}} = \frac{q\eta}{d}\sqrt{\frac{k_{\rm B}\TCM}{M}}.
\end{equation}

The scaling $I_{\mathrm{max}}\propto q/\sqrt{M}$ is favourable when working with highly charged massive particles. A silica sphere of radius $r_{\rm S} = 1\,\mu$m ($M = 5.5\times10^{12}\,$amu) and realistic charge $q = 10^6\,e$ (Appendix 2), where $e$ is the elementary charge, will induce a comparable current to the atomic ion $^{88}$Sr$^+$.

\section{Resistive cooling}  

\begin{figure}[t]
	 {\includegraphics{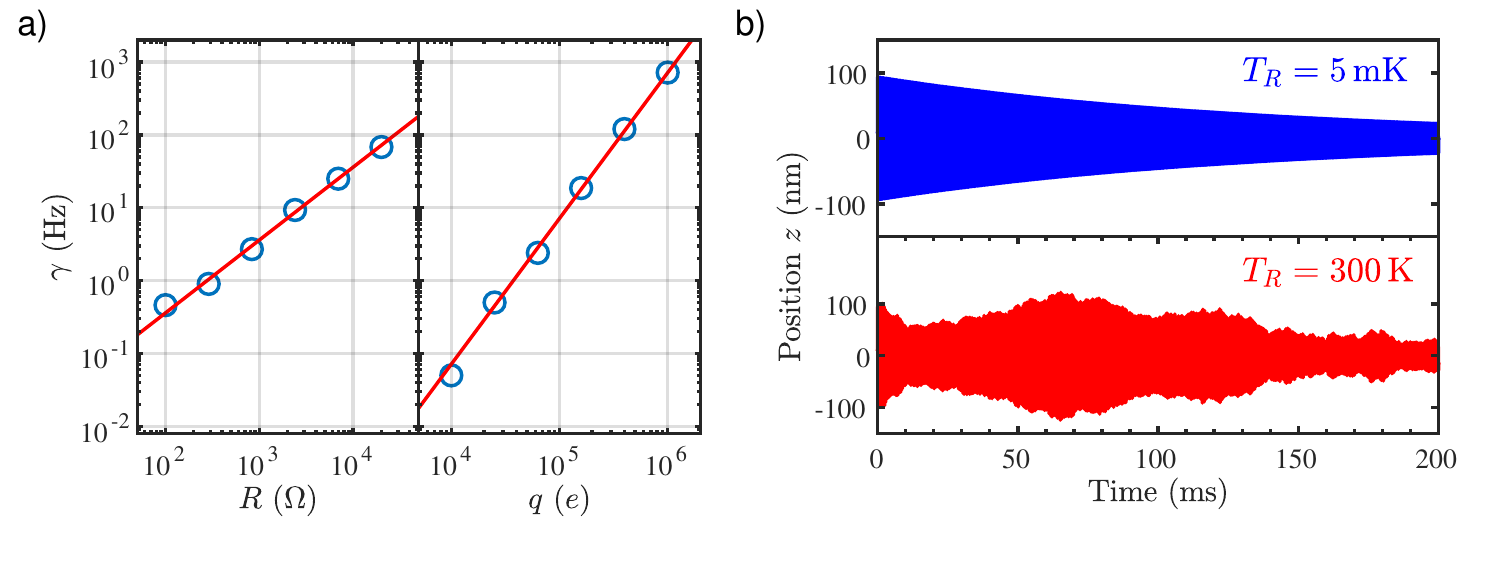}}	\centering
\caption{\label{fig:resistive} 
\textbf{Resistive cooling:} a) Variation in the momentum damping rate $\gamma$ of a charged particle coupled to a circuit, with resistor value $R$ and particle charge $q$ (open circles: simulation eqn.~\eqref{eqn:fulleom}, solid lines: eqn.~\eqref{eqn:resdamp}). b) Simulated particle trajectories for two different circuit temperatures $T_R$, illustrating that the particle thermalizes with the circuit. The parameters in this figure unless stated otherwise are: $T_{\rm in} = 1000\,$K, $r_{\rm S} = 1\,\mu$m, $q = 10^5\,e$, $d = 1\,$mm, $Q_{\rm f} = 100$, $R = 100\,$M$\Omega$, $T_R = 300\,$K.}
\end{figure}

As demonstrated in Sec. \ref{sec:dissipation}, connecting the endcap electrodes to an $RLC$ circuit serves to dissipate the induced current and thus damp the particle motion. The ensuing friction rate \eqref{eq:resdamp} on resonance ($\omega_{LC} = \omega_z$) can be written as
\begin{equation} 
\label{eqn:resdamp}
\gamma_{\rm res} = \left(\frac{q\eta}{d}\right)^2\frac{\Rf}{M},
\end{equation}
where we introduced the effective resistance $R_{\rm eff}$  \cite{Itano1995, Domizio2015}. For a series $RLC$ circuit $\Rf = Q_{\rm f}^2 R$, while for parallel it reads $\Rf = \omega_z L Q_{\rm f}$, with the circuit quality factor $Q_{\rm f} = \omega_{\rm z} / \Gamma$.

Tuned circuits with $Q_{\rm f} = 25,000$ have been used to cool N$_2^+$ ions \cite{Cornell1989}, and it is proposed to exploit high $Q_{\rm f}$ quartz crystal oscillators connected in parallel with the endcaps to further boost resistive cooling \cite{Kaltenbacher2011}. For a modest $Q_{\rm f} = 100$ and $R = 100\,$M$\Omega$, an $r_{\rm S} = 1\,\mu$m, $q = 10^6\,e$ silica sphere at room temperature would generate a signal of $\sim 100$\,mV for electrodes separated by 1\,mm.


Resistive cooling is illustrated in fig.~\ref{fig:resistive}, where the motion of a charged microsphere in a quadrupole ion trap is numerically simulated including all sources of noise, as outlined in Appendix 1. From eqn.~\ref{eqn:resdamp}, it is clear that to increase the damping rate, one can increase $\Rf$ or $q$, as verified in fig.~\ref{fig:resistive} a), where eqn.~\ref{eqn:resdamp} is compared to the results of the numerical simulation, with the agreement illustrating that realistic experimental noise has little effect on the damping rate. The scaling of the damping rate $\gamma \propto q^2/M$ is favourable for highly-charged massive particles.

The dissipation of energy across $R$ is accompanied by heating due to Johnson-Nyquist noise \cite{Brown1986, Itano1995, Domizio2015}, modeled as a white voltage-noise source $\nR$ of width $\sqrt{4k_{\rm B}T_R\Rf}$, where $T_R$ is the temperature of the circuit. For a description of how noise is added to the simulations of trapped particle motion, see Appendix 1. In the absence of other noise sources, the particle will equilibrate to the temperature of the circuit $T_R$, as indicated by the thermal trajectory in fig.~\ref{fig:resistive}b). Hence, to reach low temperatures via resistive cooling, cryogenic circuitry is required \cite{Domizio2015}. Note that it is possible to design ion trap geometries to resistively cool all three degrees-of-freedom, for example by using a ring electrode which is split into segments \cite{Itano1995, Horvath1997}.

Even under ambient conditions, resistive cooling can be useful to stabilize trapped particles, for example against collisions with residual gas molecules \cite{Kiesel2013, Millen2014, Ranjit2015}, or to pre-cool massive particles, which in general possess initial motional energies far above room temperature due the loading mechanism \cite{Millen2015, Millen2016}. Furthermore, avoiding optical fields removes motional heating due to scattering of photons. 

To consider the limits of resistive cooling, we consider operation in a state-of-the-art dilution refrigerator at 5\,mK. By considering the quadrupole ion trap stability parameters, as defined by eqn.~\eqref{eqn:stab} in Appendix 1, an $r_{\rm S} = 1\,\mu$m silica sphere with $q=10^6\,e$ can have stable frequencies of $\omega_z > 2\pi\times1\,$MHz, which would correspond to reaching a phonon occupancy of $n = k_{\rm B}\TCM/(\hbar\omega_z) < 100$. 

\section{Feedback cooling}
\label{sec:feedback}

\begin{figure}[!t]
	 {\includegraphics{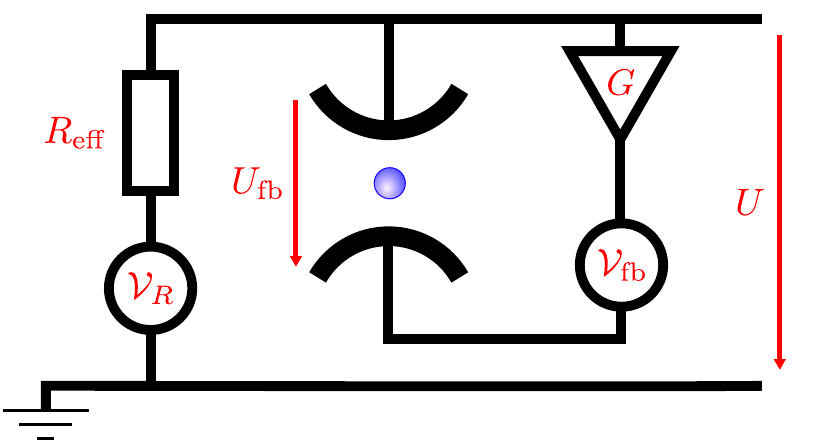}}	\centering
\caption{\label{fig:feedback} 
Circuit diagram for the feedback control of charged particle motion. The motion of the particle induces a voltage $U$ across the effective resistance $\Rf$. An amplifier of gain $G$ feeds this voltage back onto one of the endcap electrodes, effectively shorting out the Johnson-Nyquist noise at $G=1$. The feedback amplifier introduces voltage noise $\nfb$. }
\end{figure}

Since we can detect the motion of a charged particle via the endcap voltage $U$, we can feed this signal back onto the endcaps to dynamically control the particle's motion. Figure~\ref{fig:feedback} shows a schematic of this process, where an amplifier with gain $G$ generates the voltage $GU$ on the lower endcap electrode. With the appropriate phase shift between $U$ and $GU$, this either amplifies or cools the motion \cite{Dehmelt1986, DUrso2003, DUrsoThesis}. 

\begin{figure}[!t]
	 {\includegraphics{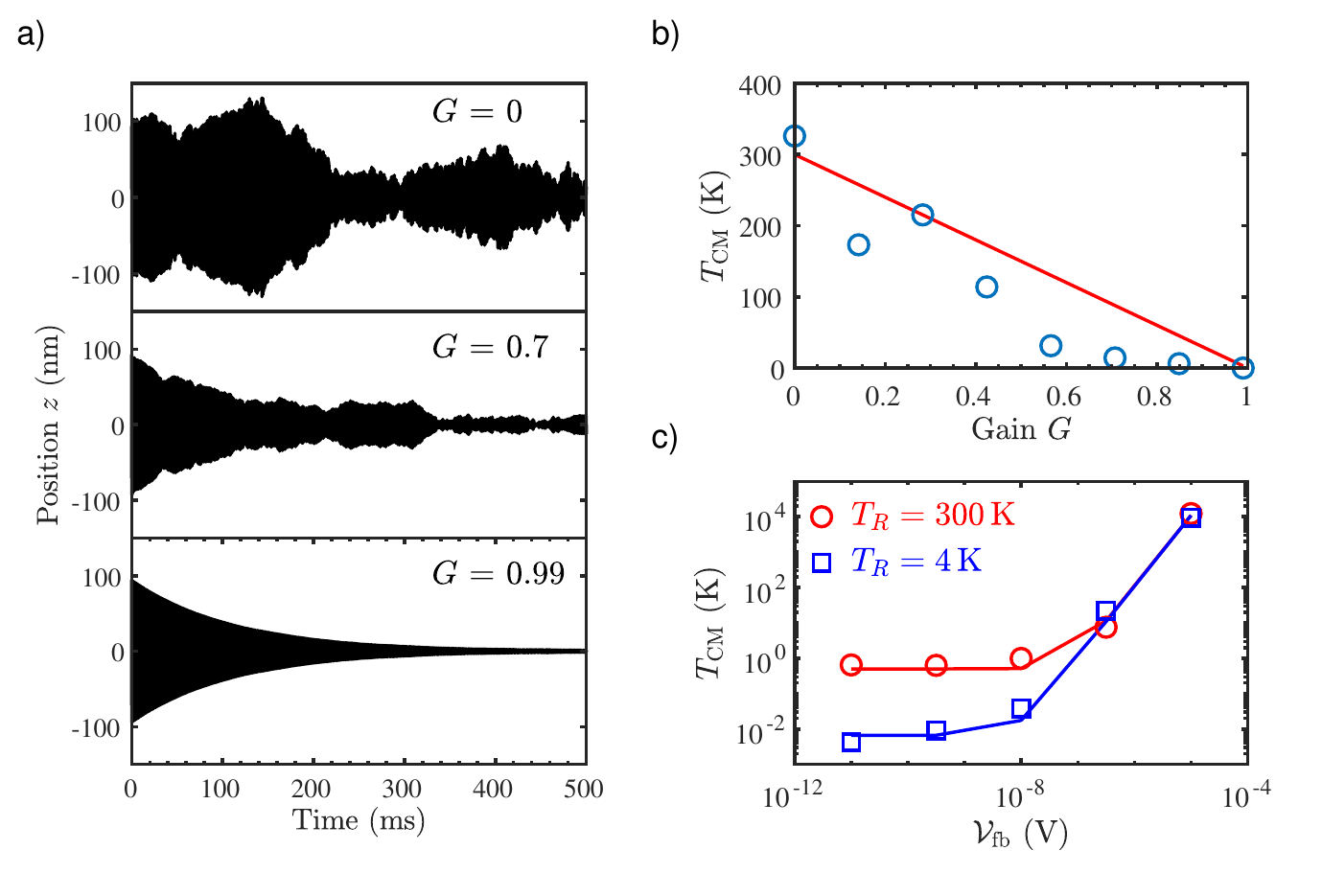}}	\centering
\caption{\label{fig:feedback_cooling} 
\textbf{Feedback cooling:} a) Simulated particle trajectories showing the effect of feedback gain $G$. Variation in simulated particle equilibrium temperature $\TCM$ with b) amplifier gain $G$ (open circles: simulation eqn.~\eqref{eqn:fulleom}, solid line: eqn.~\eqref{eqn:fbtemp}, disagreement due to the finite run-time of the simulations), and c) feedback amplifier noise voltage $\nfb$ at two different circuit temperatures $T_R$, with $G$ = 0.95 (open circles: simulation eqn.~\eqref{eqn:fulleom}, solid line: eqn.~\eqref{eqn:fbtemp}). The parameters in this figure unless stated otherwise are: $T_{\rm in} = 1000\,$K, $r_{\rm S} = 1\,\mu$m, $q = 10^5\,e$, $d = 1\,$mm, $Q_{\rm f} = 100$, $R = 100\,$M$\Omega$, $\nfb = 10^{-10}\,$V, $T_R = 300\,$K.}
\end{figure}

In the scenario depicted in fig.~\ref{fig:feedback}, the charged particle sees a voltage $U_{\mathrm{fb}}$ which is generated by the voltage difference between the two endcap electrodes: 

\begin{equation} 
\label{eqn:fb}
U_{\mathrm{fb}} = (1-G)U.
\end{equation}

At $G = 1$ the circuit resistance that the particle sees $R_{\mathrm{fb}} = (1-G)\Rf$ and the thermal noise is tuned out. Explicitly, in this simple model the temperature $\TCM$ of the particle is reduced to $\TCM = (1-G)T_R$. Simultaneously, the feedback damping rate $\gamma_{\mathrm{fb}} \propto R_{\mathrm{fb}}$ goes to zero as $G$ approaches unity, $\gamma_{\mathrm{fb}} = (1-G)\gamma$. Charged nano- and micro-particles are stable in Paul traps on the timescale of months \cite{Millen2015}, somewhat negating the necessity for rapid damping in the absence of other noise sources.

For a more realistic description, it is necessary to include the voltage noise added by the feedback amplifier $\nfb$, as illustrated in fig.~\ref{fig:feedback}. Noise added due to collisions with gas molecules and electrode surface potentials is considered in the Supplementary Information, and included in the simulations of the particle motion presented in figs.~\ref{fig:resistive} \& \ref{fig:feedback_cooling}, and can be neglected. This noise voltage has an associated noise temperature $\Tn$ defined through $\nfb = \sqrt{4k_{\rm B}\Tn R_{\mathrm{amp}}B}$, where $R_{\mathrm{amp}}$ is the resistance of the feedback amplifier, and $B$ is its bandwidth. This noise adds in quadrature with the uncorrelated voltage noise $\nR$ \cite{DUrsoThesis}, leading to a total

\begin{equation} 
\label{eqn:fbnoise}
\ntot = \sqrt{(1-G)^2\nR^2 + G^2\nfb^2},
\end{equation}

\noindent and a particle equilibrium temperature,

\begin{equation} 
\label{eqn:fbtemp}
\TCM = (1-G) T_R + \frac{G^2}{1-G}\Tn.
\end{equation}

A simulation of the effect of feedback on the particle dynamics is shown in fig.~\ref{fig:feedback_cooling}a). A comparison of eqn.~\eqref{eqn:fbtemp} and simulation is shown both for a variation in feedback gain $G$ and in noise voltage $\nfb$ in figs.~\ref{fig:feedback_cooling}b) \& c) respectively. Equation~\ref{eqn:fbtemp} is minimized at $G \approx 1-\sqrt{\Tn/T_R}$, yielding a minimum temperature $\Tmin \approx 2\sqrt{\Tn T_R}$. The noise voltage $\nfb$ of commercial amplifiers can be sub-nanoVolt, yielding sub-Kelvin particle temperatures even with room temperature circuitry, as illustrated in fig.~\ref{fig:feedback_cooling}c).

Using state-of-the-art cryogenic SQUID amplifiers with noise temperatures below $20\,\mu$K \cite{Vinante2001} with $T_{R} = 5\,$mK, it would be possible to use feedback to reduce the temperature to below $100\,\mu$K. This corresponds to the motional ground-state for few-MHz oscillation frequencies. Hence, with cryogenic operation, this system is in-principle suitable for cooling micron-sized charged particles to the quantum ground-state.

\section{Position, size and force sensitivity}

Opto- and electro-mechanical devices make excellent sensors, due to their low mass and ability to couple to a wide range of forces \cite{Metcalfe2014}, and they have found application in genetics, proteomics, microbiology and studies of DNA \cite{Tamayo12}. Their sensitivity is limited by dissipation to the environment, a problem which gets worse with decreasing physical size \cite{Ekinci2005}. By levitating the oscillator, many dissipation processes are removed, which has enabled force sensing with optically levitated microparticles on the zepto-Newton scale \cite{Ranjit2016}.

\begin{figure}[t]
	 {\includegraphics{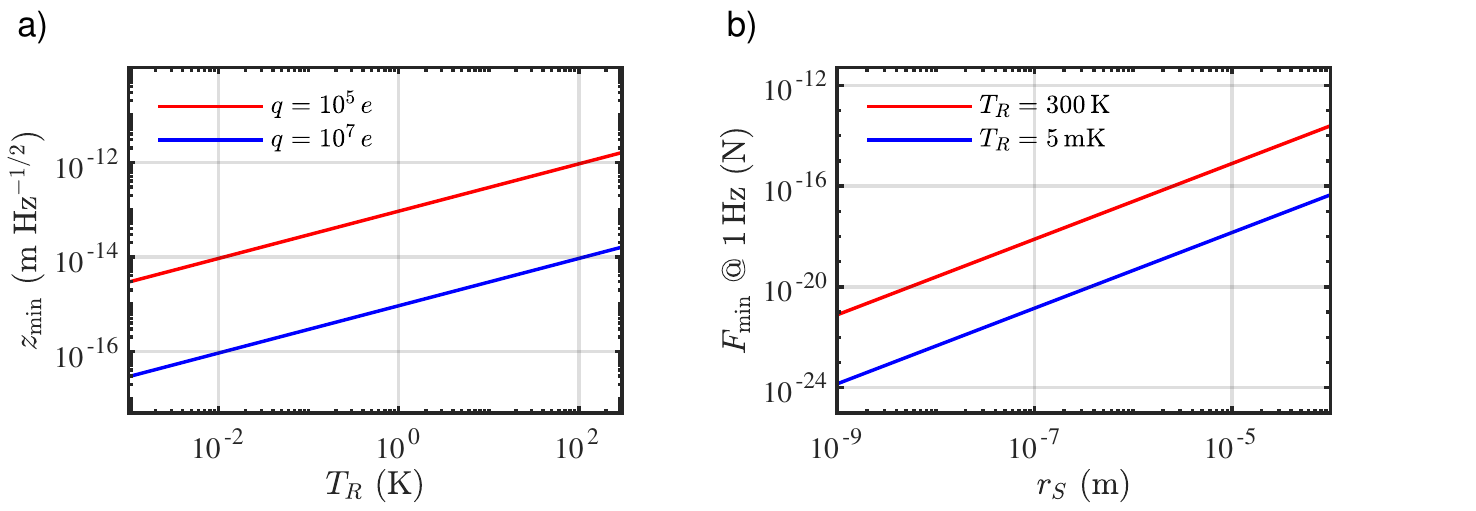}}	\centering
\caption{\label{fig:sensing} 
\textbf{Sensing:} a) Minimum detectable particle displacement with a 1\,Hz measurement bandwidth as a function of the circuit temperature $T_R$, for two different values of particle charge $q$, with $\omega_z = 2\pi\times 1\,$MHz. b) Minimum detectable force with a 1\,Hz measurement bandwidth, as a function of particle size $r_{\rm S}$ (silica sphere), for two different circuit temperatures $T_R$, assuming that the particle thermalises with the circuit. The parameters in this figure unless stated otherwise are: $r_{\rm S} = 1\,\mu$m, $d = 1\,$mm, $Q_{\rm f} = 100$, $R = 100\,$M$\Omega$.}
\end{figure}

The smallest displacement of a charged particle that can be measured through an induced voltage in this system is limited by noise. The dominant source is the Johnson-Nyquist noise voltage $\nR = \sqrt{4k_{\rm B}T_R \Rf\Delta\nu}$, where $\Delta\nu$ is the detection bandwidth. Since one can make a phase sensitive detection \cite{Brown1986}, $\Delta\nu$ can be very small. By considering the point at which the induced signal $U$ is equal to $\nR$, it is straightforward to show that the minimum detectable particle velocity on resonance is

\begin{equation}
\label{eqn:minz} 
	\dot{z}_{\mathrm{min}} = \sqrt{\frac{4k_{\rm B}T_R\Delta\nu}{M\gamma}},
\end{equation}
since other sources of noise are negligible (Supplementary Information). For a harmonic oscillator one has $z_{\rm min}=\dot{z}_{\rm min}/\omega_z$. The variation in $z_{\mathrm{min}}$ with $T_R$ and $q$ is shown in fig.~\ref{fig:sensing}a). Considering a $r_{\rm S} = 500\,$nm silica sphere, $\gamma = 1\,$kHz, and $\omega_z = 2\pi\times1\,$MHz, even at room temperature (300\,K) this corresponds to $10^{-11}\,$m\,Hz$^{-\frac{1}{2}}$ resolution, and at a cryogenic temperature of 5\,mK, $< 10^{-15}\,$m\,Hz$^{-\frac{1}{2}}$ resolution. The zero-point fluctuation of such a particle is $z_{\mathrm{zpf}} = \sqrt{\hbar/2M\omega_z} = 8\times10^{-15}\,$m. 

When considering the force sensitivity of a harmonic oscillator, the minimum detectable force is $F_{\mathrm{min}} = \sqrt{4k_{\rm B}\TCM M\gamma_{\rm CM}\Delta\nu}$, where $\gamma_{\rm CM}$ is the damping rate on its motion \cite{Ranjit2016}. In our case, the dominant damping without feedback is resistive $\gamma$, as defined in eqn.~\eqref{eqn:resdamp}. This cannot be made arbitrarily low, since small $\gamma$ means we detect less signal $U$, which must in turn be greater than the noise voltage $\nR$. Hence, we have the requirement,

\begin{equation}
\label{eqn:detlim} 
	I\Rf > \sqrt{4k_{\rm B}T_R \Rf \Delta\nu}.
\end{equation}

Using eqns.\ \eqref{eqn:maxcurrent} \& \eqref{eqn:resdamp}, and assuming that the particle is in thermal equilibrium with the circuit $\TCM = T_R$, we find the simple expression $\gamma > 4\Delta\nu$, which is the value of the resistive damping required to measure above the thermal noise in a given bandwidth. In other words, the measurement time must be on the order of the ring-down time ($1/\gamma$). This yields a force sensitivity

\begin{equation}
\label{eqn:force} 
	F_{\mathrm{min}} = \gamma\sqrt{k_{\rm B}T_RM}.
\end{equation}

The variation in $F_{\mathrm{min}}$ with particle size and temperature is shown in fig.~\ref{fig:sensing}b). Practically, choosing a low value of $\gamma$ will lead to long thermalisation times, requiring a stable experiment, although particles in ion traps are stable on month-timescales. Using eqn.~\eqref{eqn:force}, with a 1\,Hz measurement bandwidth, for an $r_{\rm S} = 100\,$nm silica sphere, $F_{\mathrm{min}}$ at 300\,K is $8\times10^{-19}\,$N, and at 5\,mK is $3\times10^{-21}\,$N, comparable to state-of-the-art optical systems \cite{Ranjit2016}. It is worth noting that it is possible to levitate lower-mass particles in an ion trap than optically, due to an increased trap depth, making the electromechanical system attractive for sensing applications.

Finally, using eqns.~\eqref{eqn:maxcurrent} \& \eqref{eqn:detlim}, it is possible to detect particles which satisfy the following relation,

\begin{equation}
\label{eqn:QoverM} 
	\frac{q}{\sqrt{M}} > \sqrt{\frac{2\Delta\nu}{\Rf}}\frac{d}{\eta}.
\end{equation}

As an example, this means that using realistic experimental parameters ($R = 100\,$M$\Omega$, $Q_{\rm f} = 100$, $\eta = 0.8$, $d = 1\,$mm, $\Delta\nu = 1\,$Hz), it would be possible to detect singly charged particles with masses up to $5\times10^6$\,amu.

\section{Discussion}

We have demonstrated that \emph{levitated electromechanics}, where charged particles levitated in an ion trap are interfaced with an $RLC$ circuit, can be used for electronic detection, cooling and precision sensing. Using feedback, sub-Kelvin temperatures are achievable with room temperature circuitry, and we predict that working in a cryogenic environment will enable ground-state cooling for micron-sized particles.

Levitated electromechanics is compatible with optomechanical systems. A hybrid levitated opto-/electro-mechanical is suitable for cooling deep into the quantum regime, and for the production of interesting mechanical quantum states, such as squeezed states \cite{Genoni2015}. Charged particles equilibrate with the trapping circuitry, which acts as a highly controllable thermal bath, making the system suitable for studies of low-dissipation thermodynamics \cite{Millen2018}. 

\section*{Acknowledgments}

JM is grateful for discussions with Andrew Geraci, and for funding from EPSRC project EP/S004777/1. DG would like to thank Peter Barker for useful discussions, and is supported by the Controlled Quantum Dynamics Centre for Doctoral Training at Imperial College London. BS acknowledges support by Deutsche Forschungsgemeinschaft (DFG -- 411042854). TN acknowledges support from Austrian Science Fund (FWF) Projects Y951-N36 and F4019-N23.

\section*{Appendix 1: Equations of motion in a quadrupole ion trap}

For the purposes of this work, we consider only motion along the $z$-axis between the endcap electrodes of a spherical quadrupole ion trap. The equation-of-motion when driven by a voltage $U_{\rm AC}(t) = U_{\mathrm{DC}} + U_0\cos(\omega_D t)$ is

\begin{equation}
M \ddot{z}  -\frac{M\omega_D^2}{4}[a_z -2q_z\cos(\omega_Dt)]z=0,
\end{equation}
where $\omega_D$ is the drive frequency, and $a_z, q_z$ are stability parameters, 

\begin{equation}
\label{eqn:stab}
\eqalign{a_z &= \frac{4U_\mathrm{DC} \eta q}{M\omega_D^2r_0^2} \cr
q_z &= -\frac{2U_0 \eta q}{M\omega_D^2r_0^2}},
\end{equation}
with $\eta=0.8$ is a geometric factor, $M$  the mass of the particle, and $2r_0$ the separation between the RF electrodes (or the diameter of the ring electrode). In what follows, we set $U_\mathrm{DC} = 0$ (and hence $a_z = 0$). This yields a secular frequency in the $z$-direction of $\omega_z = \omega_D q_z/2 \sqrt{2}$.

Feedback, as discussed in Sec.~\ref{sec:feedback}, acts as a force proportional to the velocity $\dot{z}$, since it depends on the induced current $I$ (see eqn.~\eqref{eqn:current}). Hence, with feedback, the equation-of-motion reads:

\begin{equation}
M \ddot{z} + M\gamma_{\mathrm{fb}}\dot{z} + \frac{M\omega_D^2}{2}q_z\cos(\omega_Dt)z=0.
\end{equation}

\subsection*{Inclusion of noise}

We consider the influence of several noise sources: gas collisions at temperature $T_{\rm gas}$ through the damping rate $\gamma_{\mathrm{gas}}$ (see Supplementary Information); the Johnson-Nyquist noise of the circuit at the temperature $T_{\rm R}$ through the rate $\gamma$; and the noise of the feedback amplifier $\nfb$ with associated noise temperature $\Tn$ and damping rate $\gamma_{\mathrm{fb}}$. 

We include each of these in our model via the fluctuation-dissipation theorem, 

\begin{equation}
\label{eqn:noise}
\eqalign{ \langle F_{\mathrm{gas}}(t+\tau)F_{\mathrm{gas}}(t) \rangle = 2k_{\rm B}T_{\mathrm{gas}}\gamma_{\mathrm{gas}}M\delta(\tau)\cr
 \langle F_R(t+\tau)F_R(t) \rangle = 2k_{\rm B}T_R\gamma M\delta(\tau)\cr
 \langle F_{\mathrm{fb}}^{\mathrm{n}}(t+\tau)F_{\mathrm{fb}}^{\mathrm{n}}(t) \rangle = 2k_{\rm B}\Tn \gamma_{\mathrm{fb}}M\delta(\tau),}
\end{equation} 
for the gas, resistive, and feedback random force respectively.

The electrode surface noise, which is discussed in detail in the Supplementary Information, is included by considering the fluctuating force
$\langle F_E(t+\tau)F_E(t) \rangle = q^2S_E(\omega_z)\delta(t-\tau)$, constructed noting that an electric field $E$ generates a force $qE$ on a particle of charge $q$.

Hence, the full equation-of-motion for the $z$-direction reads:

\begin{equation}
\label{eqn:fulleom}
 \eqalign{M \ddot{z} + M(\gamma_{\mathrm{gas}} + \gamma_{\mathrm{res}} + \gamma_{\mathrm{fb}})\dot{z} +\frac{M\omega_D^2}{2}q_z\cos(\omega_Dt)z 
 = F_{\mathrm{gas}} + F_{\mathrm{res}}+ F_{\mathrm{n,fb}}+ F_E.}
\end{equation}

\section*{Appendix 2: Charging the particles}
\label{sec:charging}

The resistive and feedback damping rates depend on the charge $q$ of the trapped particle. A dielectric particle of radius $r_{\rm S}$ can hold a maximal negative charge \cite{Draine1987} of about

\begin{equation} 
\frac{q_{\rm neg}}{e} = -1 -0.7\left (\frac{r_{\rm S}}{\rm nm}\right )^2,
\end{equation}
and a maximal positive charge of \cite{Draine1987}

\begin{equation}
\label{eq:charge} 
\frac{q_{\rm pos}}{e} = 1 + 21\left (\frac{r_{\rm S}}{\rm nm}\right )^2.
\end{equation}

This implies that a $r_{\rm S} = 1\,\mu$m particle can hold up to $2\times10^7$ positive charges. Practically, $5\,\mu$m diameter melamine particles have been charged via positive ion bombardment (5\,keV He ions) to hold $\sim 7 \times 10^6\,\vert e\vert$ \cite{Velyhan2004}, significantly below the theoretical limit in eq.~\eqref{eq:charge}.

Charging can be achieved via electron or ion bombardment \cite{Schlemmer2001, Velyhan2004, Kopnin2017}, corona discharge \cite{Adamiak2002} and adhesion of charged droplets. Not much is known about the charging limits for sub-micron particles, but it has been noted that, for smaller particles, secondary emission of charge from the bulk material can limit the maximum surface potential \cite{Svestka1995, Kopnin2017}. Smaller spheres require higher electric fields to charge, as determined by the Pauthenier equation \cite{Adamiak2002}, which says that in a uniform electric field $E$, the maximum charge held by a sphere is

\begin{equation} 
	q_{\rm max} = 4\pi\epsilon_0r_{\rm S}^2pE,
\end{equation}
where $p = 3$ for a conductor, and $p = 3\epsilon_{\rm r}/(\epsilon_{\rm r} +2)$ for a dielectric. Low charges on nano- \cite{Frimmer2017} and micro-particles \cite{Schlemmer2001} are stable over timescales of hours.

\newpage

\begin{center}
\textsc{\LARGE Supplementary Information}\\[1.2cm]
\end{center}

\setcounter{equation}{0}
\setcounter{figure}{0}
\setcounter{table}{0}
\setcounter{page}{1}
\makeatletter

\section*{Interaction with residual gas}
\label{sec:gas}

In discussions of cooling ions or electrons, it is assumed that the background pressure is as low as possible, since gas collisions lead to trap loss. This is not necessarily true for much more massive nano- and micro-particles, where collisions with background gas lead to thermalization with the environmental temperature $T_\mathrm{gas}$. Indeed, in nano- and micro-particle trapping experiments, this dissipation process is necessary for loading conservative trapping potentials, with loading pressures ranging from 0.1\,mbar \cite{Millen2015}, 3\,mbar \cite{Kuhn2016}, 10\,mbar \cite{Ranjit2015}, to atmospheric \cite{Kiesel2013}. Furthermore, nano- and micro-particles in optical traps are observed to be unstable below pressures ranging from $10^{-5}\,$mbar \cite{Vovrosh2017} to $\approx 1\,$mbar \cite{Kiesel2013, Millen2014, Ranjit2015}. 

\begin{figure}[t]
	 {\includegraphics[width=0.95\textwidth]{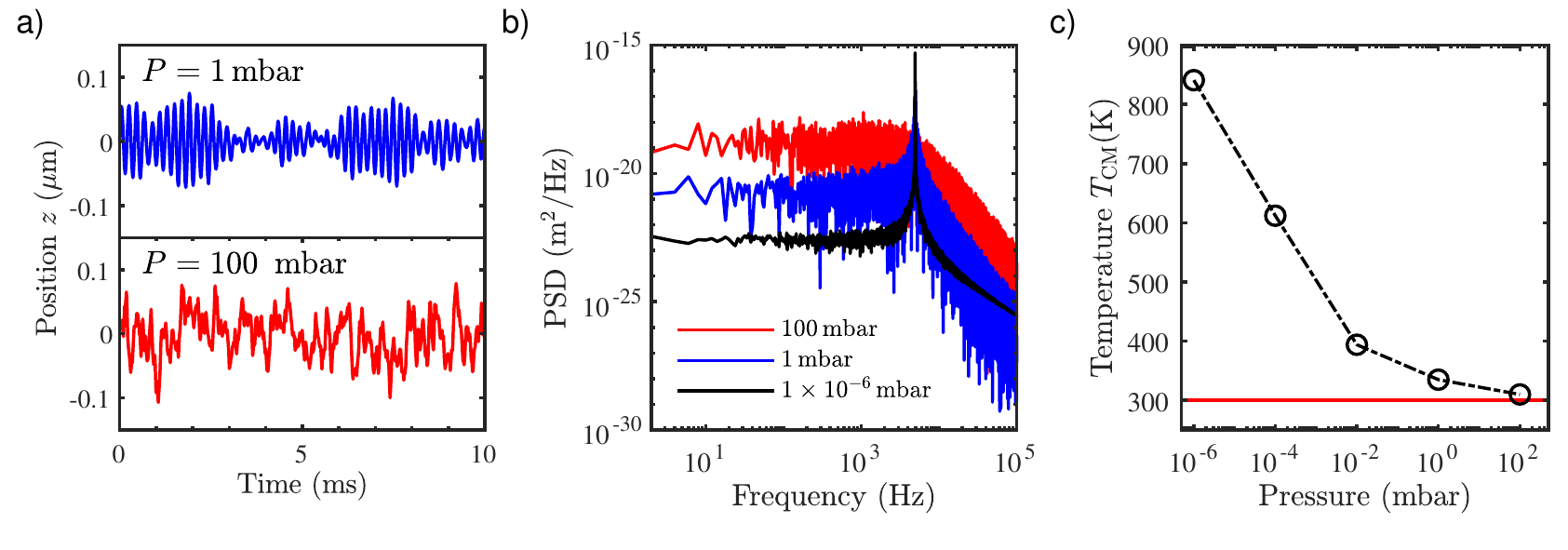}}	\centering
\caption{\label{fig:gas}
\textbf{Background gas pressure:} a) Simulated particle trajectories at two different gas pressures $P$. b) Simulated PSD of motion for different $P$. c) Variation in particle temperature $\TCM$ after 500\,ms with $P$, for an initial particle temperature $T_{\mathrm{in}} = 1000\,K$ and gas temperature $T_{\mathrm{gas}} = 300\,K$, showing that at higher pressures the particle rapidly thermalises with the gas (simulation: open circles, theory (300 K): solid line). Feedback gain $G=0$ for these simulations, $q = 10^5\,e$, $R = 100\,\Omega$, $Q_{\rm f} = 100$, $T_R = 300\,$K.}
\end{figure}

The damping rate on a sphere of radius $r_{\rm S}$ is \cite{Millen2014, Martinetz2018}:

\begin{equation} 
\label{eqn:gas} \tag{S1}
\gamma_{\mathrm{gas}} = \frac{4\pi}{3}\frac{m n_{\rm gas}r_{\rm S}^2{\bar v}_{\mathrm{th}}}{M},
\end{equation}

\noindent where $m$ is the mass of the gas molecules, $n_{\rm gas}$ is the number density of the gas, ${\bar v}_{\mathrm{th}}$ is the mean gas thermal velocity, and $M$ is the mass of the sphere. As a sense of scale, for a $r_{\rm S}=1\,\mu$m silica sphere, at a pressure of 100\,mbar (N$_2$), $\gamma_{\mathrm{gas}} \sim 10^4\,$Hz, and at $10^{-8}\,$mbar, $\gamma_{\mathrm{gas}} \sim 10^{-6}\,$Hz. This means that, when working at UHV, even modest cooling rates overcome heating due to gas collisions. Figure~\ref{fig:gas} shows simulations of the effect of residual gas on trapped particle dynamics.

\section*{Electrode surface noise}
\label{sec:EFnoise}

A major source of noise when considering atomic ions comes from both static and varying electric fields originating from electrode surfaces \cite{Brownutt2015}. Following this reference and \cite{Goldwater2016}, we express the heating rate of the trapped particle due to electric field noise $\gamma_E$ in terms of the spectral density $S_E(\omega)$ of the noise:

\begin{equation} 
\label{eqn:Enoise1} \tag{S2}
\gamma_E = \frac{q^2}{4M\hbar\omega_z}S_E(\omega).
\end{equation}

\begin{figure}[!t]
	 {\includegraphics[width=0.95\textwidth]{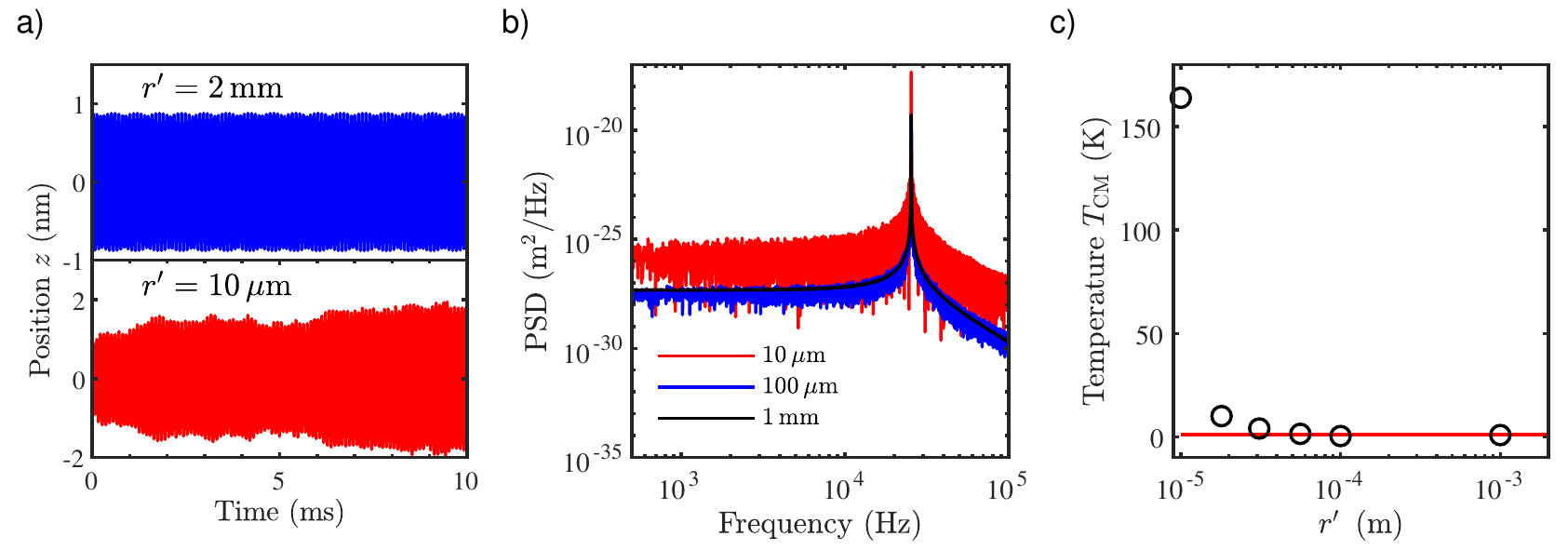}}	\centering
\caption{\label{fig:electrode_noise} 
\textbf{Electrode noise:} The effect of electrode surface noise is explored through simulation by varying the trap-centre to endcap electrode spacing $r'$. a) Simulated particle trajectories for two different values of $r'$. b) Simulated PSD of particle motion for different $r'$. c) Simulated particle temperature as $r'$ is varied, showing heating for small values of $r'$. The initial particle temperature is $T_{\mathrm{in}} = 1\,$K (red line in c)), $T_{R} = 300\,$K, $q = 10^6\,e$, $G=0$. The resistance $R = 1\,\Omega$ and $Q_{\rm f} = 100$ in these simulations, to remove resistive cooling. For all data presented in this paper, the electrode noise parameters as defined in eqn.~\ref{eqn:Enoise2} are $\alpha = 1, \chi = 2, \beta = 3, g_E = 10^{-12}$.}
\end{figure}

Values of $S_E(\omega)$ can be estimated from the literature, however it is useful to understand the scaling of this noise with experimental parameters. It is typical \cite{Brownutt2015} to treat the electric field noise as colored, with power spectral density

\begin{equation} 
\label{eqn:Enoise2} \tag{S3}
S_E = g_E \omega^{-\alpha}r_0^{\beta}T_E^{\chi},
\end{equation}
where $T_E$ is the temperature of the electrodes, which we set equal to $T_{R}$. For the purposes of this work, following the literature, we take $g_E = 10^{-12}, \alpha = 1, \chi = 2, \beta = 3$. The effect of electric field noise is shown in fig.~\ref{fig:electrode_noise}. Figure ~\ref{fig:electrode_noise} c) only shows significant heating for highly charged particles ($q=10^6\,e$) at sub-$100\ \mu$m particle-electrode distances. Previous work has noted that this is not a significant source of noise for a charged nanoparticle (as opposed to atomic ions), due to its large mass \cite{Goldwater2016}, and this work confirms this viewpoint.

\section*{Simulation parameters}

In this manuscript, we always consider a quadrupole ion trap with $r_0 = 500\,\mu$m unless otherwise stated. Table~\ref{table:params} lists the ion trap operating parameters used in each figure. The pressure is always held at $10^{-10}$\,mbar unless otherwise stated, though the dynamics are not effected at any pressure below $10^{-6}$\,mbar, and the background gas is always considered to be at $300\,K$.

\begin{table}[!t]
	\centering
		\begin{tabular}{ |l|c|c|c|c|c|c|c|c|c|c|c|c| }
		\hline
			Figure &  $V_0$ (V) & $f_D$ (MHz) \\
			\hline 
			2 a-b) 				& 3000 & 100k  \\
			\hline
			4 a-c) & 3000 & 100k  \\
			\hline
			\ref{fig:gas} a)-c) 					& 3000 & 200k  \\
			\hline
			\ref{fig:electrode_noise} a)-c) & 1000 & 100k \\
			\hline

		\end{tabular}
		\caption{\label{table:params} Quadrupole ion trap parameters used in the simulated data for the figures.}
\end{table}

\end{document}